\documentclass[sigconf,nonacm]{acmart}
\AtBeginDocument{
  }

\setcopyright{none}
\begin{document}

\title[Feasibility-Aware Latent Spaces for Procedural Design]{Learning Feasibility-Aware Latent Spaces for Preference-Based Exploration of Procedural Automotive Wheel Designs}

\author{Takashi Owaki}
\affiliation{
  \institution{Toyota Central R\&D Labs., Inc.}
  \country{Japan}}
\author{Yuki Koyama}
\affiliation{
  \institution{National Institute of Advanced Industrial Science and Technology (AIST)}
  \country{Japan}}
\author{Tomoyasu Nakano}
\affiliation{
  \institution{National Institute of Advanced Industrial Science and Technology (AIST)}
  \country{Japan}}
\author{Takahiro Yamaguchi}
\affiliation{
  \institution{Toyota Central R\&D Labs., Inc.}
  \country{Japan}}
\author{Masataka Goto}
\affiliation{
  \institution{National Institute of Advanced Industrial Science and Technology (AIST)}
  \country{Japan}}
\author{Hiroyuki Sakai}
\affiliation{
  \institution{Toyota Central R\&D Labs., Inc.}
  \country{Japan}}

\renewcommand{\shortauthors}{Takashi Owaki, et al.}

\begin{abstract}
Intelligent design interfaces that rely on preference-based optimization are most useful when their suggestions are both meaningful to users and feasible within the target domain. Procedural models offer compact and editable design spaces, but their native parameters can be entangled and can generate many invalid outputs, causing human-in-the-loop optimizers to waste comparisons. We propose an interaction-oriented representation-learning pipeline for procedural models and study it in automotive wheel design. The method first screens procedurally generated samples using geometric rules and finite-element analysis, then learns a reduced latent space from the screened subset. We further introduce supervised functional alignment, which reserves selected latent dimensions for stiffness, strength-related stress response, or weight so that search can be biased toward functionally meaningful regions. Simulation experiments show that screened reduction improves target-shape retrieval and the feasibility rate of suggestions, whereas unscreened reduction degrades both. Additional simulations show that constraining search along learned functional dimensions accelerates exploration toward target functional properties. A controlled study with 40 participants further shows that a 5D feasibility-aware space yields higher shape similarity and more feasible suggestions than the original 9D procedural parameterization. These results suggest that, for intelligent user interfaces in engineering design, the representation exposed to the user is a central part of the interaction design, not merely a preprocessing step for the optimizer.
\end{abstract}

\begin{teaserfigure}
  \includegraphics[width=\textwidth]{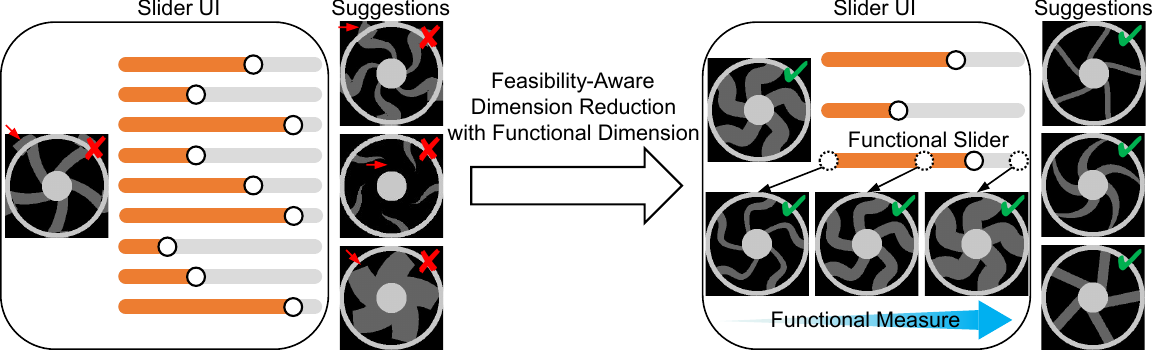}
  \caption{
  Concept of feasibility-aware latent-space learning for procedural modeling.
Green check marks and red crosses indicate feasible and infeasible designs, respectively.
A procedurally defined design space with raw parameters (left) often contains many infeasible designs, such as the left wheels shown here, which are infeasible due to protrusions and gaps (red arrows).
The design space is first restricted to feasible regions and then reduced to a smaller set of interaction dimensions with a higher likelihood of producing feasible designs (right).
The selected latent dimensions are further aligned with functional measures, enabling efficient exploration while satisfying functional requirements.
}
  \Description{Overview of feasibility-aware dimension reduction for wheel design exploration.}
  \label{fig:teaser}
\end{teaserfigure}

\maketitle

\section{Introduction}

Intelligent user interfaces increasingly combine machine-generated design alternatives with human judgment. Preference-based optimization is a representative example: rather than asking users to define a complete objective function, the system iteratively proposes candidates and learns from pairwise or set-wise choices \cite{takagi2001interactive,chu2005preference,brochu2007active,gonzalez2017preferential,koyama2017sequential,koyama2022bo}. This paradigm is well suited to design problems in which aesthetic or experiential goals are hard to formalize. Its effectiveness, however, depends not only on the optimizer but also on the representation over which the optimizer searches. If the search space contains many invalid candidates or if small parameter changes produce unintuitive visual changes, the system can repeatedly ask users to compare alternatives that are not useful for the design task.

Procedural models make this representation problem especially visible. They provide compact, editable generators of large design spaces and have been used for interactive graphics, fabrication, and suggestion-oriented design tools \cite{marks2023design,parish2001procedural,lipp2008interactive,talton2011metropolis,yumer2015procedural,shugrina2015fab}. Yet the native procedural parameters are rarely an ideal interface for human-in-the-loop search. Parameters can be entangled, Euclidean distance in parameter space may not match visual similarity, and a large portion of the combinatorial space may produce geometrically or structurally infeasible outputs. In engineering-oriented procedural design, infeasibility is not a minor nuisance: every invalid suggestion consumes user attention and weakens the feedback signal used by the optimizer.

We address this problem as \emph{interaction-oriented representation design}. Instead of treating feasibility only as an online constraint or a penalty in an acquisition function, we reshape the space before interaction begins. The proposed pipeline first screens procedural samples with domain-specific geometric and structural tests, then learns a reduced latent space from the screened subset. We also introduce supervised functional alignment, in which a selected latent coordinate is trained to track an engineering measure such as stiffness, stress response, or weight. The goal is not to claim that all latent coordinates become semantically disentangled; rather, the goal is to expose a compact space in which the optimizer is more likely to propose feasible designs and at least one coordinate can support a meaningful functional restriction.

We instantiate the approach in a procedural model of automotive wheel spokes. This application is a useful testbed because wheel design combines visual shape, structural response, and material usage. At the same time, our model is intentionally simplified: it is designed to isolate the interaction problem of learning a feasible search space, not to certify real automotive components. We therefore evaluate the approach with controlled target-retrieval simulations and a 40-participant user study. Target retrieval provides an objective proxy for whether users can guide PBO through a candidate space efficiently; it does not by itself measure open-ended creative agency or industrial design quality, which we discuss as limitations and future work.

This paper makes three contributions. First, we present a feasibility-aware latent-space learning pipeline for procedural models that combines offline screening, image-based shape features, and VAE-based dimensionality reduction. Second, we provide a reproducible way to add supervised functional dimensions to the reduced space and clarify how this differs from unsupervised dimensionality reduction. Third, we show through simulations and a human study that the learned interaction space improves target-search performance and substantially increases the rate of feasible suggestions compared with searching in the original procedural parameterization.

\section{Related Work}

\subsection{Procedural and Suggestion-Oriented Design Spaces}

Procedural modeling has long been used to generate large structured design spaces, from cities and architecture to other graphics assets \cite{parish2001procedural,lipp2008interactive,talton2011metropolis}. Design Galleries showed that exposing diverse parameter settings can help users navigate complex parameterized systems \cite{marks2023design}. For interactive creative tools, later work learned lower-dimensional or example-based interfaces that better match human exploration, including learned procedural spaces \cite{yumer2015procedural}, font exploration \cite{odonovan2014exploratory}, and layout suggestion systems \cite{odonovan2015designscape}.
Fab Forms further showed that precomputing valid regions of a low-parameter fabrication design space can support novice customization while preserving manufacturability \cite{shugrina2015fab}. Our work shares the same concern for validity, but learns a reduced interaction space from a screened procedural dataset rather than caching validity over a small exposed parameter space.

\subsection{Preference-Based and Constrained Bayesian Optimization}

Preference-driven optimization methods replace explicit scalar objectives with judgments from a user. Interactive evolutionary computation established this general paradigm \cite{takagi2001interactive}. Gaussian-process models later enabled statistically grounded preference learning and active query selection from pairwise comparisons \cite{chu2005preference,brochu2007active,houlsby2012collaborative}. PBO formalized BO under pairwise feedback \cite{gonzalez2017preferential}, and HCI systems have shown its usefulness for design assistance and visual tuning \cite{koyama2017sequential,koyama2022bo}.
Recent HCI work has further improved sample efficiency by transferring information across users and tasks, including meta-BO for wrist-based interaction calibration \cite{liao2024meta}, continual optimization across users \cite{liao2025continual}, and Meta-PO for visual appearance tuning \cite{li2025efficient}. These methods primarily improve the optimizer or its priors over fixed task parameterizations rather than learning a new global interaction space, although Meta-PO presents candidates on a locally constructed 2D search plane at each iteration \cite{li2025efficient}. Chan et al. found that BO-guided design helped novice designers explore more of the design space and reach better solutions, but reduced perceived agency and expressiveness \cite{chan2022investigating}. 
Constrained BO extends BO to settings where feasibility must also be modeled \cite{gardner2014bayesian,gelbart2014bayesian}, and Constrained PBO (CPBO) carries this idea to preference-based interaction by jointly addressing subjective objectives and inequality constraints \cite{iwai2025constrained}. Our work is complementary: instead of only learning constraints during optimization, we reshape the interaction domain offline so that the optimizer begins from a representation with a much higher prior probability of yielding feasible designs.

\subsection{High-Dimensional BO and Representation Learning}

High-dimensional black-box optimization remains difficult. Existing BO approaches address this challenge via random embeddings \cite{wang2016rembo}, local trust-region models \cite{eriksson2019scalable}, or improved priors that make vanilla BO more robust in higher dimensions \cite{hvarfner2024vanilla}. These methods are algorithmic responses to a hard search space. Our strategy is different: we change the space itself through learned reduction. Autoencoders and VAEs learn low-dimensional codes that support reconstruction, interpolation, and generation \cite{hinton2006reducing,bengio2013representation,kingma2014auto}, while learned 3D shape spaces show that latent variables can enable meaningful editing operations \cite{achlioptas2018learning}. Yumer et al. specifically demonstrated that autoencoders can create more intuitive interaction spaces for procedural models \cite{yumer2015procedural}. We build on this insight, but train the latent space only on screened feasible samples and explicitly reserve dimensions for engineering properties.

\section{Feasibility-Aware Latent-Space Learning}

\subsection{Overview}
\label{sec:method-overview}

\begin{figure*}
    \centering
    \includegraphics[width=\linewidth]{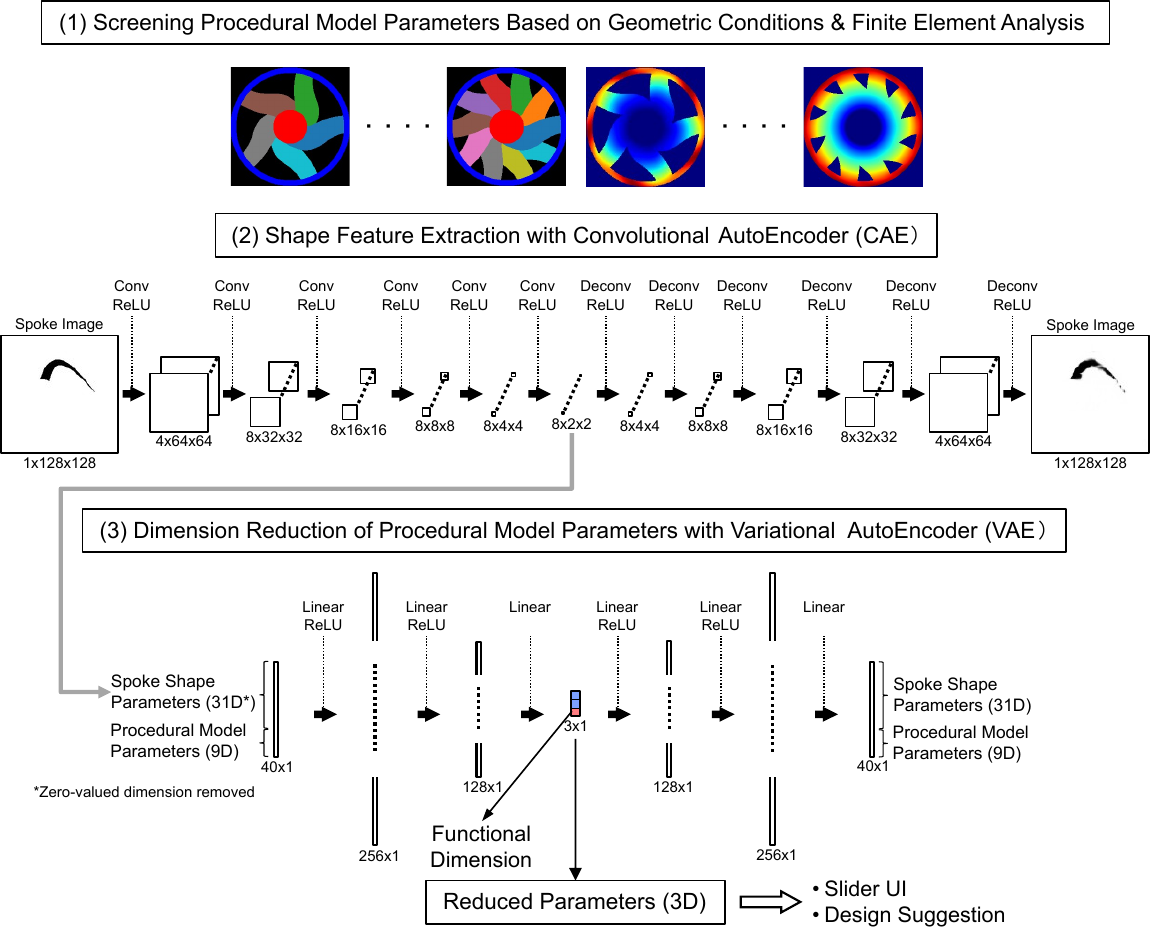}
    \caption{
Our feasibility-aware latent-space learning pipeline.
(1) Procedural parameters are screened using geometric conditions based on a single spoke and finite element analysis (FEA) with 5 to 10 spokes. The left images show wheel geometries, while the right heatmaps visualize displacement distributions from FEA; Each spoke, rim, and hub is colored differently for clarity.
(2) A convolutional autoencoder extracts spoke-shape features from single-spoke images.
(3) A variational autoencoder (VAE) learns reduced parameters from the concatenated procedural and shape features, with a selected latent dimension aligned to a functional measure. Although separate VAEs are trained for 3D, 4D, and 5D latent spaces, that for the 3D latent space is shown here.      }
    \Description{Schematic illustration of the proposed feasibility-aware latent space learning pipeline.}
    \label{fig:feas-aware-dim-reduct}
\end{figure*}

Figure~\ref{fig:feas-aware-dim-reduct} summarizes the pipeline. Let $\Theta$ denote the full discrete set of procedural parameter combinations and let $\Theta_{\mathrm{scr}}\subset\Theta$ denote the subset that passes screening. For each $\boldsymbol{\theta}\in\Theta_{\mathrm{scr}}$, we render a single-spoke image, compute functional measures from FEA, and construct a feature vector $\mathbf{x}=[\boldsymbol{\theta};\mathbf{h}]$ by concatenating the original parameters with a learned spoke-shape descriptor $\mathbf{h}$. A variational autoencoder (VAE) then maps $\mathbf{x}$ to an $m$-dimensional latent code $\mathbf{z}$, where $m\in\{3,4,5\}$. PBO operates in the normalized latent space as well as in the original procedural domain. When functional alignment is enabled, one latent dimension is trained to track a normalized engineering measure, allowing the interface to bias or restrict search along a functionally meaningful axis.

This design has two motivations. First, screening removes large invalid regions before any user interaction occurs; in our dataset only 30,502 of 2,450,000 parameter combinations survive screening. Second, combining procedural parameters with image-based spoke features encourages neighborhoods in the reduced space to reflect both parametric and geometric similarity, which is useful for slider-based exploration.

\subsection{Procedural Wheel Model}
\label{sec:proc-model}

To demonstrate the method, we built a procedural wheel model in Blender Geometry Nodes. The model is deliberately compact rather than industrially complete: it lets us study whether a feasibility-aware representation improves interactive search while keeping the design space small enough to enumerate, screen, and inspect. A single spoke is defined by nine parameters (Table~\ref{tab:proc-parameters}), including the locations of the two internal B\'{e}zier control points, a planar translation, an in-plane rotation, and the slope and intercept that control the long side of a rectangular cross-section. A full wheel is formed by replicating this spoke radially and combining it with a fixed hub (radius 0.06~m) and rim (inner radius 0.19~m, outer radius 0.21~m).

\begin{figure}[tb]
  \centering
  \includegraphics[width=0.7\linewidth]{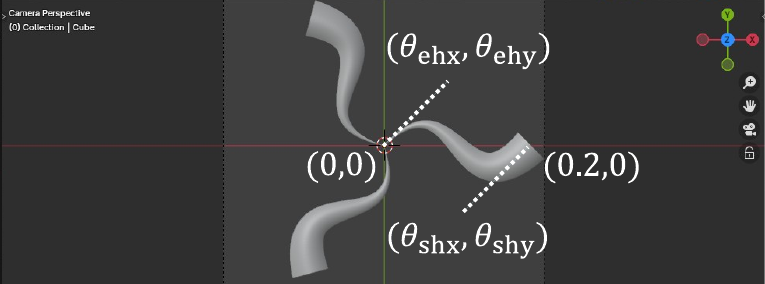}
  \caption{
  Illustration of procedural model parameters related to the B\'{e}zier curve.
  The wheel geometry is constructed by radially repeating a single spoke (three-spoke example shown) and combining it with a fixed hub and rim (not shown).
  }
  \label{fig:spoke-shape}
  \Description{A curved spoke with labeled Bezier control-point parameters.}
\end{figure}

The center line of the spoke is a B\'{e}zier curve with endpoints at $(0,0)$ and $(0.2,0)$~m, and the spoke body is modeled as a solid with a rectangular cross-section. Figure~\ref{fig:spoke-shape} illustrates the control-point parameters. The discrete values listed in Table~\ref{tab:proc-parameters} are exhaustively enumerated to create the dataset used for screening and representation learning, yielding 2,450,000 parameter combinations. This enumeration is a design choice for the present study, not a requirement of the general pipeline; larger procedural grammars could use offline sampling or adaptive coverage of the feasible region, as discussed in Section~\ref{sec:discussion}.

\begin{table*}
\centering
\caption{Parameters for Procedural Wheel Model}
\label{tab:proc-parameters}
\begin{tabular}{lcl}
\toprule
Description & Parameter & Values Used for Dimension Reduction \\
\midrule
Start control point ($x$) of B\'{e}zier curve [m]
& $\theta_{\mathrm{shx}}$
& $0.02 \times \{1, 4, 7, 10\}$ \\
Start control point ($y$) of B\'{e}zier curve [m]
& $\theta_{\mathrm{shy}}$
& $0.02 \times \{-8, -4, 0, 4, 8\}$ \\
End control point ($x$) of B\'{e}zier curve [m]
& $\theta_{\mathrm{ehx}}$
& $0.02 \times \{1, 4, 7, 10\}$ \\
End control point ($y$) of B\'{e}zier curve [m]
& $\theta_{\mathrm{ehy}}$
& $0.02 \times \{-8, -4, 0, 4, 8\}$ \\
Translation along $x$-axis [m]
& $\theta_{\mathrm{tx}}$
& $0.005 \times \{-4, -2, 0, 2, 4\}$ \\
Translation along $y$-axis [m]
& $\theta_{\mathrm{ty}}$
& $0.005 \times \{-4, -2, 0, 2, 4\}$ \\
Rotation angle about start point [rad]
& $\theta_{\mathrm{rz}}$
& $0.05 \times \{-4, -2, 0, 2, 4\}$ \\
Slope of rectangular cross-section long side [m]
& $\theta_{\mathrm{m}}$
& $0.01 \times \{-12, -8, -4, 0, 4, 8, 12\}$ \\
Intercept of rectangular cross-section long side [m]
& $\theta_{\mathrm{a}}$
& $0.01 \times \{0, 2, 4, 6, 8, 10, 12\}$ \\
\bottomrule
\end{tabular}

\footnotesize
The long side length of the rectangular cross section is defined as
$\theta_m t + \theta_a$, where $t \in [0,1]$ denotes the normalized
position along the B\'{e}zier curve.
\end{table*}

\subsection{Screening Procedural Parameters}
\label{sec:screening}

The screening stage removes designs that are unsuitable for interactive exploration. We use a two-level test: geometric validity ensures that a parameter set yields a meaningful wheel geometry, and FEA-based structural screening removes cases that do not remain stable under our simplified loading scenario. Figure~\ref{fig:feas-aware-dim-reduct}(1) summarizes this process.

\paragraph{Geometric screening.}
We apply four checks: (1) both ends of a spoke must lie inside the hub or rim; (2) the spoke width must remain non-negative; (3) the spoke must not extend beyond the outer edge of the rim; and (4) the B\'{e}zier center line must not self-intersect. Samples that satisfy these conditions are referred to as \emph{feasible shapes}. These checks eliminate obviously invalid geometries that would otherwise consume user attention during PBO.

\paragraph{Structural screening.}
Geometric validity does not guarantee that a design is suitable for exploration, so we additionally approximate structural behavior with FEA. We fix the wheel at the hub and apply a uniform pressure of 1~MPa toward the wheel center on the outer rim, which serves as a simplified proxy for loading during early-stage exploration. It is not intended to model the full automotive validation regime: cornering loads, impact loads, fatigue, thermal effects, manufacturing tolerances, and regulatory safety margins are outside the scope of this study. We evaluate each spoke geometry in wheel configurations with 5 to 10 spokes using FEniCS, a mesh size of 0.005~m, and an aluminum alloy A5052 material model (Young's modulus $70\times 10^9$~Pa, Poisson ratio 0.33). A parameter set is retained only if displacement and von Mises stress can be computed successfully for every tested spoke count. This criterion is conservative for the interaction study because it removes numerically unstable or structurally pathological cases before they can be shown to users.

After screening, 30,502 parameter sets remain out of 2,450,000 candidates. The large reduction confirms that the feasible region is sparse and motivates learning the interaction space directly from the screened subset rather than from the full raw domain.

\subsection{Spoke-Shape Feature Extraction with a Convolutional Autoencoder}
\label{sec:cae}

The nine procedural parameters do not fully capture how close two wheel shapes look from a user's perspective. Small parametric changes can lead to visually disproportionate differences, and conversely distant parameter vectors can still produce similar silhouettes. To complement the raw parameters, we learn an image-based descriptor of a single spoke with a convolutional autoencoder (CAE) (Figure~\ref{fig:feas-aware-dim-reduct}(2)). Training on single-spoke images makes the descriptor independent of the number of radial repetitions in the full wheel.

The CAE contains six convolutional and six deconvolutional hidden layers with ReLU activations. It is trained on screened samples using the spoke image as both input and target output, with a batch size of 24, a learning rate of 0.001, the Adam optimizer, and 10,000 epochs. The resulting bottleneck feature vector captures local spoke geometry and is concatenated with the procedural parameter vector before the next stage.

\subsection{Latent-Space Learning with a Variational Autoencoder}
\label{sec:vae}

We learn the reduced interaction space with a VAE (Figure~\ref{fig:feas-aware-dim-reduct}(3)). For each screened sample, let $\boldsymbol{\theta}\in\mathbb{R}^9$ be the procedural parameter vector and let $\mathbf{h}$ be the CAE feature vector after removing dimensions that are always zero in the training set. We form the input vector $\mathbf{x}=[\boldsymbol{\theta};\mathbf{h}]$. The encoder outputs a diagonal Gaussian,
\begin{equation}
q_{\phi}(\mathbf{z}\mid \mathbf{x}) = \mathcal{N}\!\left(\boldsymbol{\mu}_{\phi}(\mathbf{x}),\operatorname{diag}(\boldsymbol{\sigma}_{\phi}^{2}(\mathbf{x}))\right),
\end{equation}
and the decoder reconstructs $\hat{\mathbf{x}}$ from $\mathbf{z}$. We use a VAE rather than a plain autoencoder because the KL term encourages a smoother latent distribution, which is advantageous for continuous slider-based exploration.

The network has six fully connected hidden layers with ReLU activations. We train separate models for latent dimensionalities $m\in\{3,4,5\}$ and for four alignment conditions: no functional alignment, stiffness alignment, strength alignment, and weight alignment. The total loss is
\begin{equation}
\mathcal{L}_{\mathrm{total}}
=
\mathcal{L}_{\mathrm{rec}}
+
\beta D_{\mathrm{KL}}\!\left(q_{\phi}(\mathbf{z}\mid \mathbf{x})\,\|\,\mathcal{N}(\mathbf{0},\mathbf{I})\right)
+
\gamma \mathcal{L}_{\mathrm{func}},
\label{eq:vae-total-loss}
\end{equation}
where the reconstruction loss is
\begin{equation}
\mathcal{L}_{\mathrm{rec}} = \lVert \mathbf{x}-\hat{\mathbf{x}}\rVert_2^2.
\end{equation}
We set $\beta = 0.1, \gamma = 0.2$.
When one latent dimension $z_k$ is aligned with a normalized functional measure $\tilde{y}$, we use
\begin{equation}
\mathcal{L}_{\mathrm{func}} = \left(z_k-\tilde{y}\right)^2;
\label{eq:functional-loss}
\end{equation}
for the no-alignment condition, we set $\mathcal{L}_{\mathrm{func}}=0$. At inference time, we use the encoder mean $\boldsymbol{\mu}_{\phi}(\mathbf{x})$ as the deterministic reduced representation.

Training uses a batch size of 64, a learning rate of 0.0001, the Adam optimizer, and 20,000 epochs; we keep the checkpoint with the minimum total loss. Although VAEs are often used as probabilistic generative models, we use the encoder mean after training as the coordinate for interaction.

\paragraph{Functional-measure selection.}
We selected the functional measures through a reproducible early-stage design protocol. A candidate measure should (1) correspond to a trade-off that a designer may plausibly want to control, (2) be computable automatically for every screened sample with the same offline pipeline, (3) vary sufficiently across the feasible subset to support range restriction, and (4) avoid application-specific pass/fail thresholds that would require detailed industrial requirements. Under this protocol we use three measures. \emph{Stiffness} is the inverse of the maximum displacement within the wheel. \emph{Strength-related stress response} is represented by the maximum von Mises stress under the same simplified load; lower stress corresponds to a larger structural margin, but we keep the shorter label ``strength'' in figures for readability. \emph{Weight} is approximated by the spoke-area ratio in pixels within the annular region enclosed by the rim and the hub. In another procedural domain, the same protocol would replace these labels with domain-appropriate scalar measures such as manufacturability score, energy use, or comfort.

Because Eq.~\ref{eq:functional-loss} uses explicit labels, the aligned models are supervised. We therefore separate two effects in the evaluation: the no-alignment condition tests feasibility-aware dimensionality reduction alone, whereas the alignment conditions test the additional benefit of reserving one coordinate for a chosen functional measure. The method does not claim that all latent coordinates are inherently interpretable. Instead, it makes one coordinate intentionally interpretable while retaining the remaining coordinates for shape variation. After training, each latent dimension is normalized to zero mean and unit variance over the screened training set, and the search range is set to $[-2,2]$ per dimension.

Table~\ref{tab:loss} reports the final total loss for all 12 settings. The 5D models achieve the lowest loss across all alignment conditions, so we use 5D in the main evaluation and report the 3D and 4D functional-dimension results in the appendix.

\begin{table}
  \caption{Total Loss of VAE after Training}
  \label{tab:loss}
  \begin{tabular}{ccc}
    \toprule
    Number of Dimensions&Function&Total Loss\\
    \midrule
    3 & --        & 0.923\\
    3 & Stiffness & 0.975\\
    3 & Strength  & 0.985\\
    3 & Weight    & 0.959\\
    \midrule
    4 & --        & 0.912\\
    4 & Stiffness & 0.956\\
    4 & Strength  & 0.948\\
    4 & Weight    & 0.938\\
    \midrule
    5 & --        & 0.910\\
    5 & Stiffness & 0.943\\
    5 & Strength  & 0.935\\
    5 & Weight    & 0.932\\
  \bottomrule
\end{tabular}
\end{table}

\begin{figure*}
    \centering
    \includegraphics[width=\linewidth]{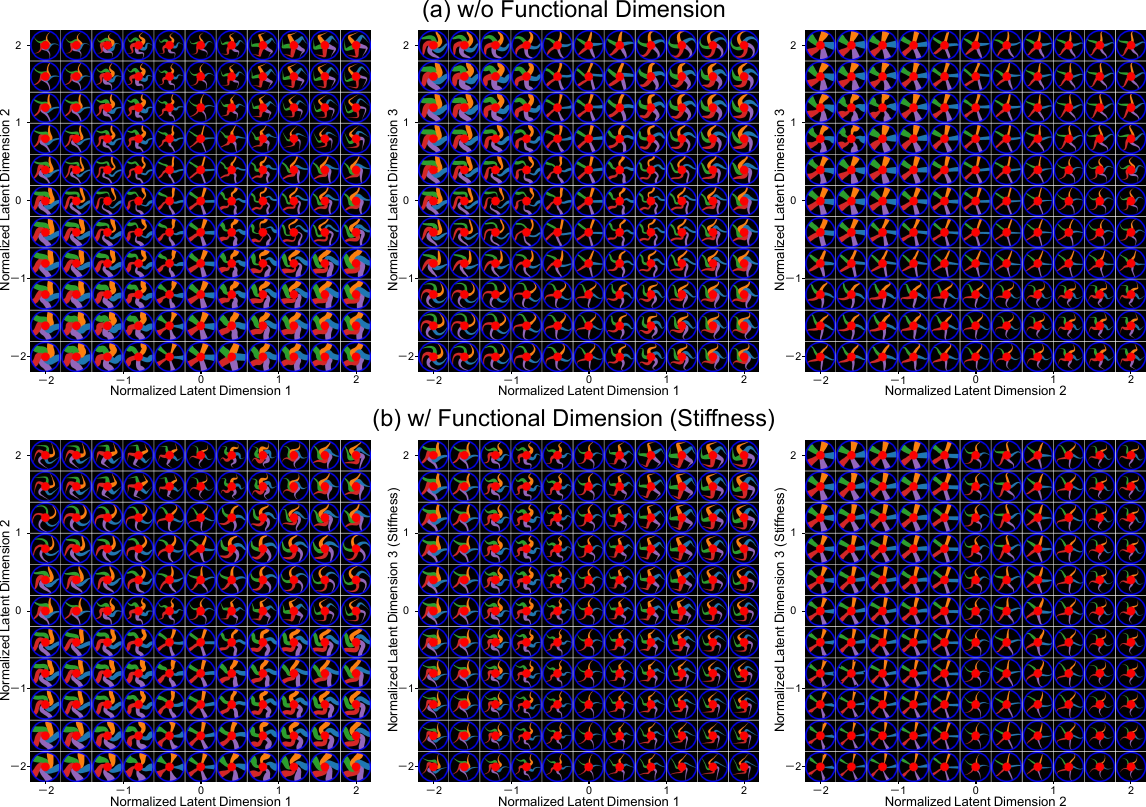}
    \caption{
Distribution of wheel shapes projected onto three planes of a 3D latent space (a) without a functional dimension and (b) with the third dimension aligned with stiffness.
Each spoke, rim, and hub is colored differently for clarity.
In both cases, the reduced space supports continuous shape variation.
With functional alignment in (b), the third axis more clearly corresponds to variations in spoke thickness that influence stiffness.
    }
    \Description{Distribution of wheel shapes on three planes in 3D latent space.}
    \label{fig:dist_shape_stiffness}
\end{figure*}

Figure~\ref{fig:dist_shape_stiffness} visualizes slices of the learned 3D spaces without functional alignment and with the third dimension aligned to stiffness. In both cases, the reduced space supports continuous shape variation. With alignment, however, the third axis more clearly corresponds to changes in spoke thickness and curvature that influence structural response, making the latent coordinate easier to interpret as a control.

\subsection{Preference-Based Exploration}
\label{sec:pbo}

Design suggestions are generated with PBO. Following the asynchronous assistant framework of Koyama and Goto \cite{koyama2022bo}, we use a Mat\'ern 5/2 kernel \cite{snoek2012practical}, the GP-UCB acquisition function \cite{srinivas2012information}, and batch candidate generation based on the parallel BO strategy of Schonlau et al.~\cite{schonlau1998global} on BoTorch \cite{balandat2020botorch}. At each iteration the system presents four candidate wheels, the user selects the design that best matches the current goal, and the observed preference is incorporated into the probabilistic model that proposes the next candidates. We run the same optimizer either in the original 9D procedural space or in the learned latent space, adjusting kernel hyperparameters to the dimensionality of the domain following Hvarfner et al.~\cite{hvarfner2024vanilla}.

This design emphasizes a different intervention point from constrained PBO (CPBO) \cite{iwai2025constrained}.
CPBO keeps the original domain and learns to respect inequality constraints during preference optimization, whereas our approach changes the domain itself before interaction by screening and compressing it. The two ideas are complementary, but our focus here is on representation design for interactive exploration.

\section{Simulation Experiments}

\subsection{Effect of Screening on Exploration Efficiency}

Our first simulation isolates the contribution of screening. The task is target retrieval: a simulated user repeatedly sees four candidate designs and selects the one whose spoke silhouette has the highest intersection-over-union (IoU) with a target design. This oracle removes human noise and lets us compare search spaces directly. We evaluate three domains: the original 9D procedural space, a screened latent space learned from the feasible subset, and a \emph{vanilla} latent space learned with the same CAE+VAE pipeline but without screening.

\begin{figure*}
    \centering
    \includegraphics[width=\linewidth]{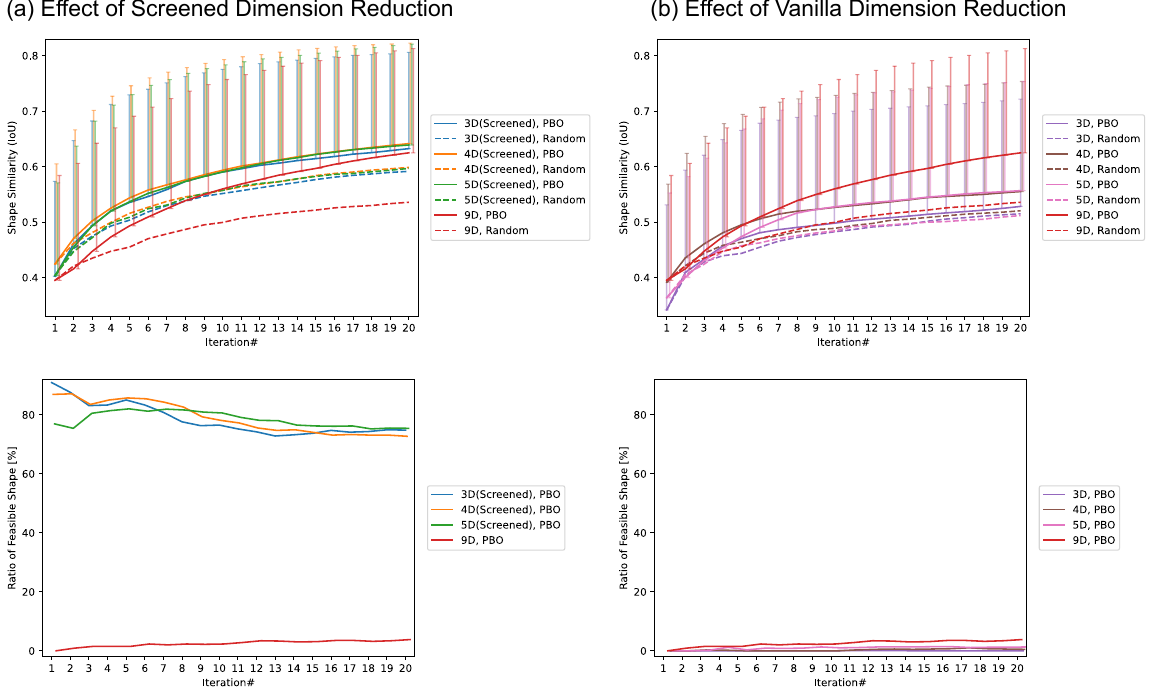}
    \caption{
    Effect of screening on preference-based optimization (PBO) without a functional dimension.
Top: Mean IoU over 20 selections for PBO and random-proposal baselines
(higher is better). For the PBO curves, error bars extend one standard
deviation above the mean; the corresponding lower error bars are omitted
to reduce visual clutter from overlapping curves.
Bottom: Mean ratio of feasible shapes among selected candidates over 20 selections for PBO (higher is better).
Screened reduction improves both search efficiency and feasibility, whereas vanilla reduction degrades both.
    }
    \Description{Plots comparing screened and vanilla dimension reduction.}
    \label{fig:effect_screening}
\end{figure*}

For each of 100 target designs sampled from the screened dataset, we run 10 trials with different random seeds and 20 preference iterations per trial. Figure~\ref{fig:effect_screening} shows mean IoU and the ratio of feasible shapes over time.
In all domains, PBO outperforms random proposal generation, confirming that preference modeling is useful. The key comparison, however, is between the latent spaces. Screened reduction improves both target-search efficiency and proposal feasibility relative to the original 9D space, whereas vanilla reduction degrades both. The result indicates that dimensionality reduction is not beneficial by itself; it helps only when the representation is learned from the region of the design space that users can actually exploit.

\subsection{Effect of Functional Dimensions on Exploration Efficiency}

We next evaluate whether functional dimensions can steer search toward target physical properties. Because the 5D models achieve the lowest training loss (Table~\ref{tab:loss}) and are used in the user study, we report the 5D results in the main paper and place the 3D and 4D results in the appendix.

\begin{figure*}
    \centering
    \includegraphics[width=\linewidth]{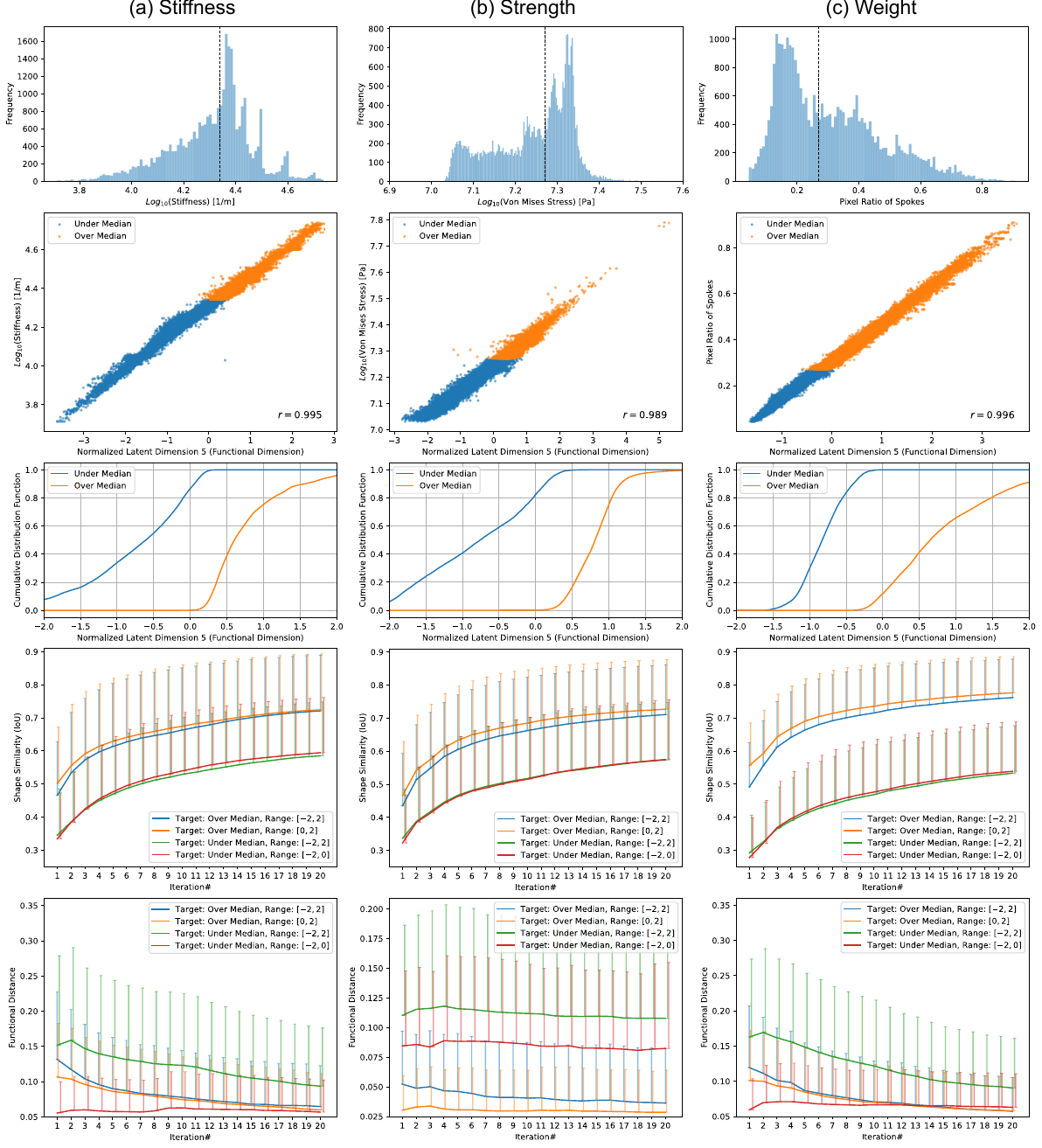}
    \caption{
    Evaluation of functional dimensions for stiffness, strength, and weight (5D).
For each functional measure, the following visualizations are shown from top to bottom: a histogram of the functional measure for the training data (the dotted line indicates the median); a scatter plot of the functional measure versus the learned functional dimension for the training data; the cumulative distribution of the training data along the learned functional dimension; the mean IoU over 20 selections (higher is better); and the mean functional distance over 20 selections (lower is better).
In the bottom two rows, error bars extend one standard deviation above
the mean; the corresponding lower error bars are omitted to reduce visual
clutter from overlapping curves.
}
    \Description{Per-property plots showing learned functional dimensions and their effect on optimization.}
    \label{fig:effect_funcdim}
\end{figure*}

For each property (stiffness, strength, or weight), we first verify that the designated latent coordinate tracks the normalized property value. Figure~\ref{fig:effect_funcdim} shows strong monotonic relations between the learned coordinate and the target measure, and the cumulative distributions indicate that negative coordinate values mostly correspond to below-median properties while positive values mostly correspond to above-median properties. This suggests a simple interaction strategy: when the user wants a lower-than-median value for a property, the system can restrict the coordinate to $[-2,0]$ instead of the full range $[-2,2]$; for higher-than-median values, it can restrict the coordinate to $[0,2]$.

We test this idea with the same target-retrieval simulation as above. For each property, we draw 100 target designs and run 10 trials per target. If the target property is below the median, we compare unrestricted search with search restricted to $[-2,0]$; if it is above the median, we compare unrestricted search with search restricted to $[0,2]$. Figure~\ref{fig:effect_funcdim} shows that range restriction consistently increases IoU and reduces the distance to the target property. The functional dimension therefore does more than improve interpretability: it provides a simple control that can bias PBO toward regions that are simultaneously more shape-relevant and more property-consistent.

\section{User Study}

Simulation shows that screening and functional alignment can help an oracle user. We next test whether the screened latent space also improves real human exploration. The task mirrors the simulation setting: participants see a target wheel and repeatedly choose, from four suggestions, the design that best matches it. We compare the original 9D procedural space with the screened 5D latent space, because 5D provided the best reconstruction loss and a practical number of interaction dimensions.

The study is intentionally controlled. Target retrieval gives every participant the same objective and allows us to measure suggestion quality with IoU and feasibility rate. It is therefore a test of whether the representation helps users steer PBO through a candidate space, not a complete evaluation of open-ended creative exploration, agency, or satisfaction. We return to these missing dimensions in Section~\ref{sec:discussion}.

\subsection{Participants and Procedure}

Forty participants (S1--40, 15 male and 25 female; age $22.4 \pm 2.9$ years) took part in the study. We used a within-subjects design. Odd-numbered participants completed the two conditions in the order 5D then 9D, while even-numbered participants experienced the reverse order. For each condition, participants searched for 10 target designs sampled from the screened dataset. Each target involved 20 rounds of selection, yielding 200 choices per participant and condition. The order of the 10 targets was counterbalanced with a Williams design.

Figure~\ref{fig:user-study-gui-results}(a) shows representative initial screens for the 5D and 9D conditions. The two conditions differ only in the optimization domain. In the 9D condition, the system searches the original procedural parameters, so candidate sets may still include wheels that violate the geometric validity checks. In the 5D condition, search takes place in the screened latent space, which substantially raises the prior likelihood that a candidate is feasible. The comparison is therefore a direct test of whether the learned representation improves interactive search.

\subsection{Ethics Statement}

This study involved human participants and was approved by the Institutional Review Board of Toyota Central R\&D Labs., Inc. (Approval No. 25-37) prior to participant recruitment.
All participants provided informed consent before participating in the study.
Participation was voluntary, and all data were anonymized prior to analysis.

\subsection{Results}

\begin{figure*}
    \centering
    \includegraphics[width=\linewidth]{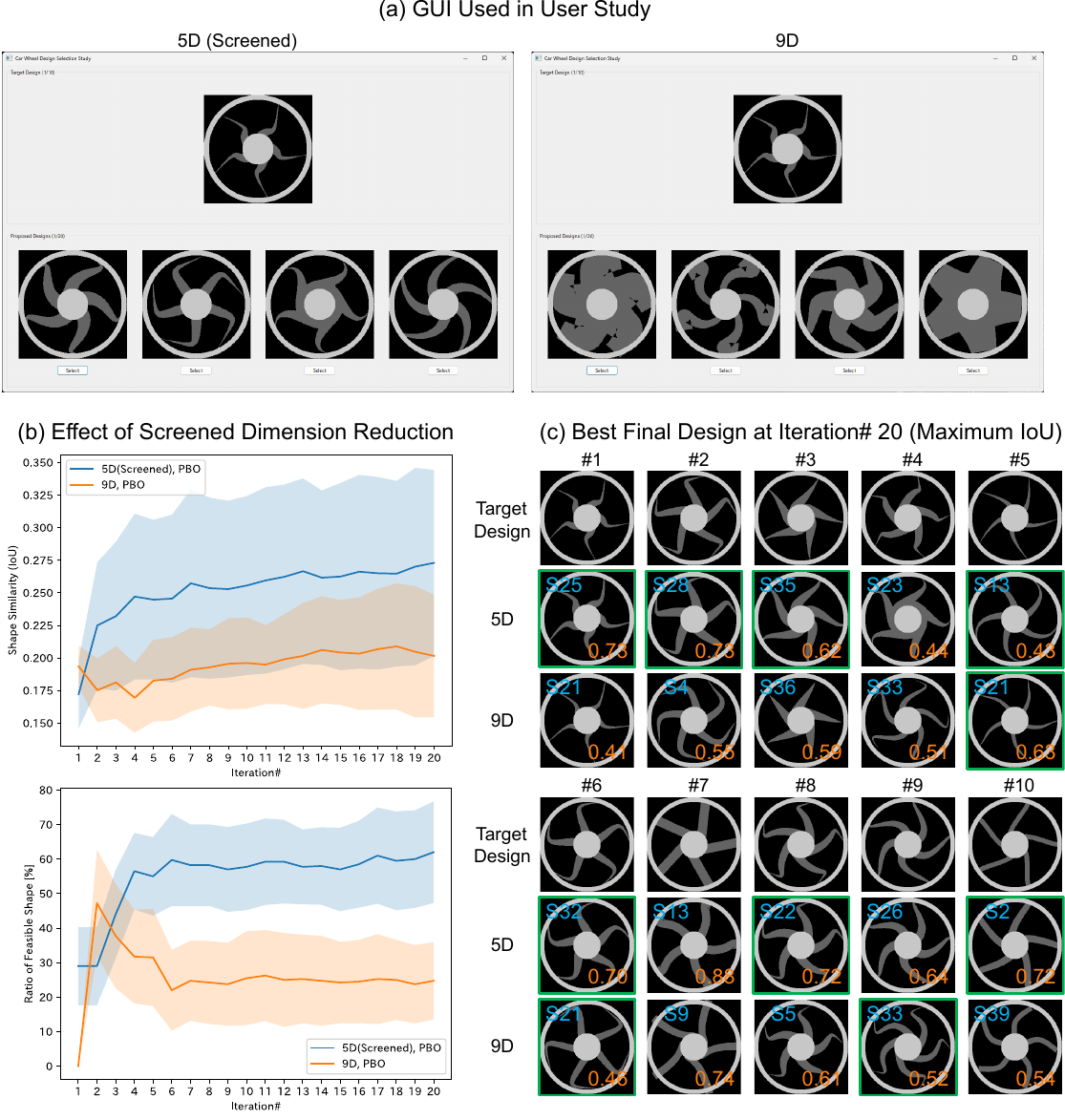}
    \caption{
(a) Example initial GUIs for the 5D (left) and 9D (right) conditions.
(b) Evolution of mean IoU (±SD; higher is better) and the ratio of feasible shapes (±SD; higher is better) over 20 selections.
(c) Best final designs at iteration\# 20 based on maximum IoU. Blue text indicates subject IDs, and orange numbers indicate IoU values. Designs enclosed by green lines indicate feasible shapes.
    }
    \label{fig:user-study-gui-results}
    \Description{Result of user study.}
\end{figure*}

Figure~\ref{fig:user-study-gui-results}(b) plots the mean IoU and mean feasible-shape ratio over the 20 selections. For each participant, we first averaged the results over the 10 targets and then computed across-participant means and standard deviations. The 5D condition outperforms the 9D condition on both measures for nearly the entire interaction. After 20 selections, paired $t$-tests on participants' mean IoU and feasible-shape ratio show significant differences ($p = 7.4 \times 10^{-7}$ and $p = 3.8 \times 10^{-15}$, respectively). These results replicate the simulation findings with human users: a feasibility-aware latent space makes each comparison more informative because fewer suggestions are wasted on implausible designs.

Figure~\ref{fig:user-study-gui-results}(c) summarizes the best design reached after 20 selections for each target. The 5D condition yields higher IoU for 8 of the 10 targets and produces feasible finalists more often. Response intervals were also slightly shorter in the 5D condition ($5.47 \pm 1.52$~s vs.\ $5.82 \pm 1.46$~s; paired $t$-test, $p = 0.0164$), suggesting a modest reduction in cognitive effort. We interpret this timing result cautiously, but it is consistent with the qualitative observation that the 5D condition presents fewer obviously invalid candidates.
Nine subjects (S2, S13, S22, S23, S25, S26, S28, S32, and S35) and seven subjects (S4, S5, S9, S21, S33, S36, S39) reached at least one best final design for 5D and 9D conditions, respectively.
There are no overlaps in the two subject groups, which suggests individual suitability differences between the two conditions.

\section{Discussion}
\label{sec:discussion}

\subsection{Interaction Contribution and Evaluation Scope}

The results suggest that the main contribution of the method is not compression alone but \emph{interaction-oriented representation design}. When the latent space is trained on all procedural samples, invalid and semantically unhelpful regions are still encoded, so PBO continues to spend comparisons on candidates the user cannot meaningfully use. Screening changes the problem before optimization begins. The resulting latent space is organized around designs that survive domain-specific validity tests, which explains why screened reduction improves both IoU and feasible-shape ratio while unscreened reduction does not.

At the same time, our evaluation should not be read as evidence that the system already supports all aspects of creative design practice. The target-retrieval task was useful because it gave us a ground-truth measure for comparing search spaces, but real designers often work with evolving, subjective, and underspecified goals. Future IUI studies should therefore include open-ended tasks in which participants formulate their own design briefs, revise goals over time, and report perceived agency, satisfaction, trust, and serendipitous discovery. Such studies would complement the present evidence rather than replace it: the present study establishes that the candidate space itself is easier to search under a controlled objective.

\subsection{Functional Alignment, Supervision, and Interpretability}

Functional dimensions add a second layer of utility. Standard latent variables are often hard to interpret, especially in engineering exploration where users care about trade-offs such as stiffness, stress response, and material usage. By aligning one latent coordinate with a functional measure, we create a control that users can understand and that the system can exploit to bias search. This is a supervised design choice, not an emergent property of the VAE. The contribution is therefore best described as a combination of feasibility-aware reduction and supervised functional alignment.

This clarification also limits the interpretability claim. Without the functional loss and range restriction, the latent dimensions may remain abstract, and users may not know which slider corresponds to which design consequence. The aligned coordinate mitigates this issue for one selected measure at a time, but it does not solve general disentanglement. A practical interface should therefore present functional labels, distributions, and example designs for the aligned coordinate, and should avoid implying that unaligned axes have stable semantic meanings.

\subsection{Scalability and Engineering Scope}

The current procedural wheel model and FEA setup are intentionally simplified. They omit many factors needed for real automotive engineering, including complex spoke surfaces, manufacturing constraints, dynamic loads, fatigue, thermal effects, and regulatory validation. The 1~MPa pressure load should be understood as a consistent proxy for screening and labeling designs, not as a claim of physical certification. This scope is sufficient for testing the interaction hypothesis, but deployment in professional engineering would require richer geometry, more realistic simulations, and validation by domain experts.

The pipeline also raises scalability questions. Exhaustive enumeration was practical for our nine-parameter grid, but larger procedural grammars may contain hundreds of parameters, continuous ranges, or disconnected feasible manifolds. In such cases, the offline screening stage could use stratified sampling, adaptive sampling near feasibility boundaries, surrogate models, or parallel simulation rather than exhaustive enumeration. The learned latent space would also need diagnostics for coverage and smoothness; if the feasible manifold is highly disconnected, a single Euclidean latent space may be less appropriate than multiple local spaces or a hierarchical interface. These are limitations of the present implementation, not fundamental barriers to the representation-design idea.

\subsection{Relation to Constrained Preference Optimization}

Our method is complementary to constrained preferential BO \cite{iwai2025constrained}. CPBO keeps the original domain and learns to respect constraints during preference optimization, whereas our approach changes the interaction domain before the optimization loop starts. The two approaches could be combined: feasibility-aware representation learning could raise the prior probability that suggestions are useful, while online constrained preference optimization could handle residual constraints, personalized constraints, or constraints that are too expensive to evaluate offline.

\section{Practical and Societal Impact}

Feasibility-aware representations can reduce wasted user effort in intelligent design tools by decreasing the number of invalid or obviously unusable candidates shown during interaction. This may make human-in-the-loop optimization more accessible to users who are not experts in optimization, procedural modeling, or structural simulation. At the same time, hiding infeasible regions can also hide assumptions. If the screening criteria are incomplete or biased toward a narrow notion of validity, the interface may prematurely exclude unconventional but valuable designs. Deployed systems should therefore expose the meaning and limitations of screening criteria, allow experts to audit excluded regions, and communicate that functional labels are approximations derived from a particular simulation setup.

The automotive-wheel case also illustrates a broader safety issue. A design suggested by an intelligent interface should not be treated as validated simply because it passed a simplified screening pipeline. Professional deployment would require domain-specific review, certified simulation and testing procedures, and clear responsibility boundaries between tool builders, designers, and engineers. These considerations are especially important as generative and optimization-based interfaces become easier to use in safety-relevant domains.

\section{Conclusion}

We presented a feasibility-aware latent-space learning pipeline for preference-based exploration of procedural wheel designs. The method screens procedural samples with geometric checks and simplified FEA, learns a reduced latent space from the screened subset, and optionally adds supervised functional dimensions for stiffness, strength-related stress response, and weight. Across simulation experiments and a 40-participant controlled study, the reduced space improves target-retrieval efficiency and increases the proportion of feasible suggestions compared with the original 9D procedural parameterization. More broadly, the results suggest that for intelligent user interfaces in engineering design, the quality of the interaction space is as important as the optimization algorithm operating within it.

\bibliographystyle{ACM-Reference-Format}
\bibliography{base}

\appendix

\section{Distribution of Wheel Shapes in Encoded 3D Space with Functional Dimension Aligned with Other Than Stiffness}

\begin{figure*}
    \centering
    \includegraphics[width=\linewidth]{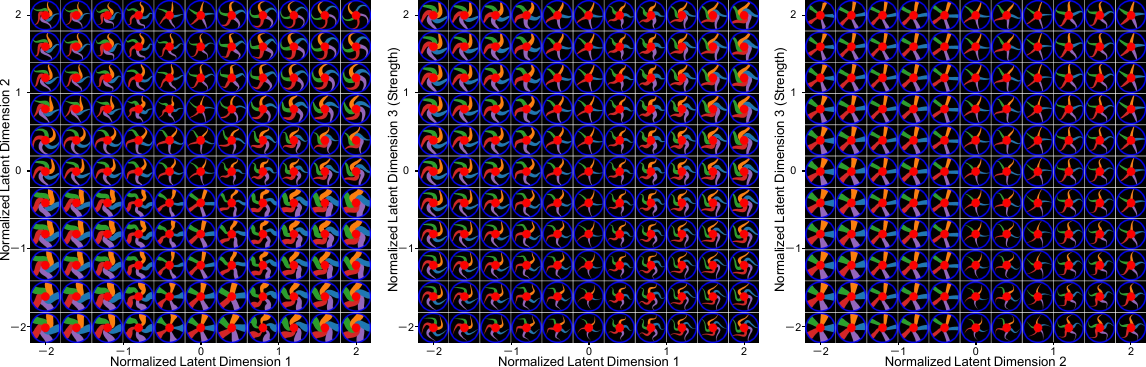}
    \caption{
    Distribution of wheel shapes on three planes of a 3D latent space with the functional dimension aligned with strength.
Each spoke, rim, and hub is colored differently for clarity.
    }
    \Description{Distribution of wheel shapes on three planes in encoded 3D space aligned with strength.}
    \label{fig:dist_shape_strength}
\end{figure*}

\begin{figure*}
    \centering
    \includegraphics[width=\linewidth]{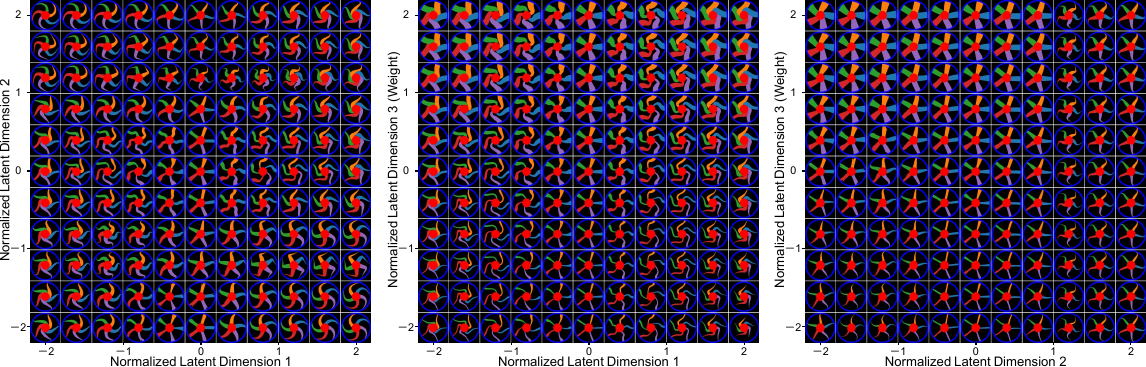}
    \caption{
    Distribution of wheel shapes on three planes of a 3D latent space with the functional dimension aligned with weight.
Each spoke, rim, and hub is colored differently for clarity.
    }
    \Description{Distribution of wheel shapes on three planes in encoded 3D space aligned with weight.}
    \label{fig:dist_shape_weight}
\end{figure*}

Figures~\ref{fig:dist_shape_strength} and \ref{fig:dist_shape_weight} show the distribution of wheel shapes on three planes in a 3D latent space when the functional dimension is aligned with strength and weight, respectively. As in Figure~\ref{fig:dist_shape_stiffness}, wheel shapes generally vary continuously on these planes, and the dominant changes along the aligned axis follow the corresponding functional measure. Because maximum von Mises stress depends on detailed local structure, some arrangements along the strength-aligned axis may appear less intuitive than those for stiffness or weight.

\section{Evaluation of Functional Dimensions for Stiffness, Strength, and Weight (3D and 4D)}

\begin{figure*}
    \centering
    \includegraphics[width=\linewidth]{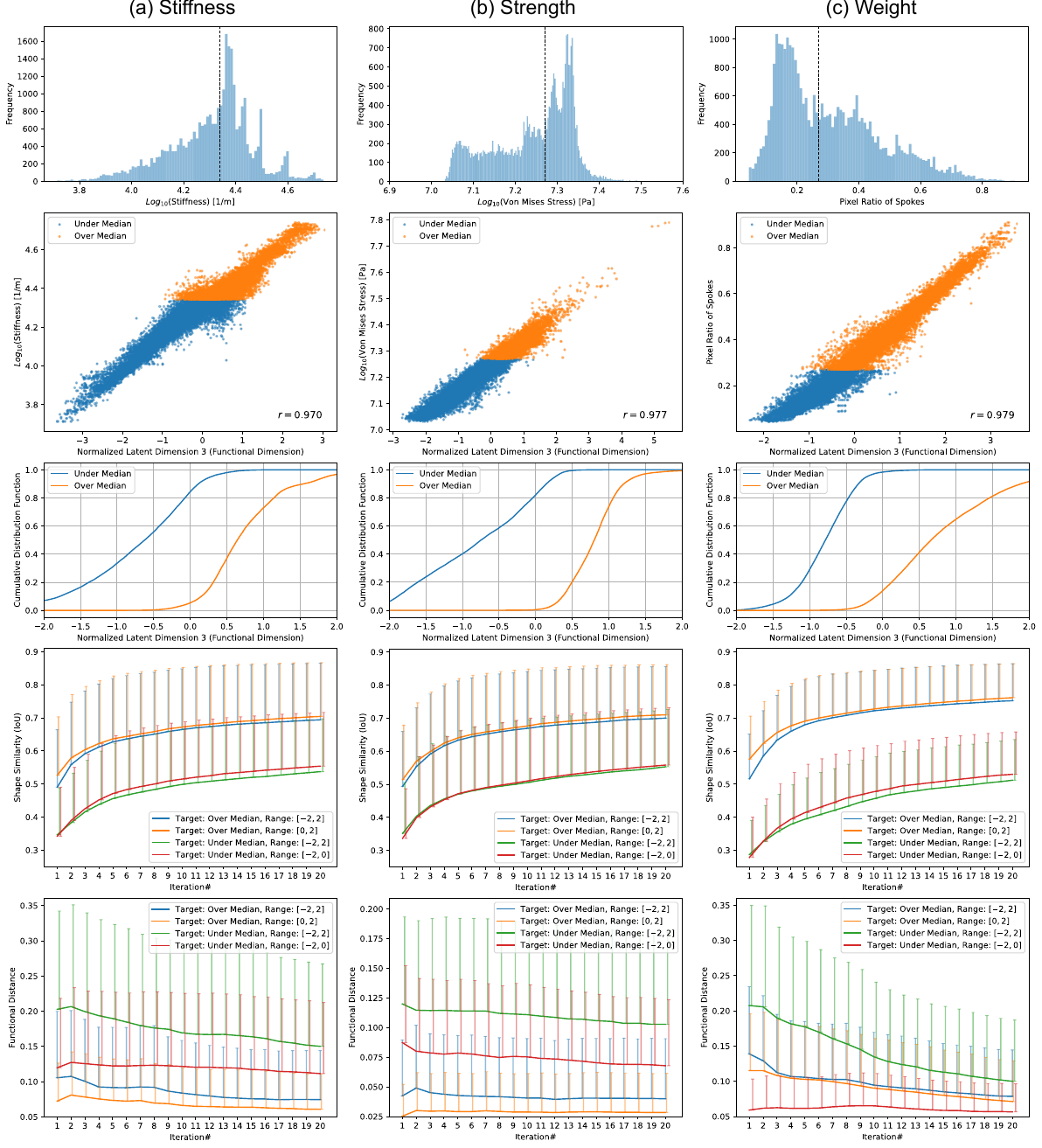}
    \caption{
    Evaluation of functional dimensions for stiffness, strength, and weight (3D).
For each functional measure, the following visualizations are shown from top to bottom: a histogram of the functional measure for the training data (the dotted line indicates the median); a scatter plot of the functional measure versus the learned functional dimension for the training data; the cumulative distribution of the training data along the learned functional dimension; the mean IoU over 20 selections (higher is better); and the mean functional distance over 20 selections (lower is better).
In the bottom two rows, error bars extend one standard deviation above
the mean; the corresponding lower error bars are omitted to reduce visual
clutter from overlapping curves.
   }
    \Description{Per-property plots showing learned functional dimensions and their effect on optimization.}
    \label{fig:effect_funcdim3}
\end{figure*}

\begin{figure*}
    \centering
    \includegraphics[width=\linewidth]{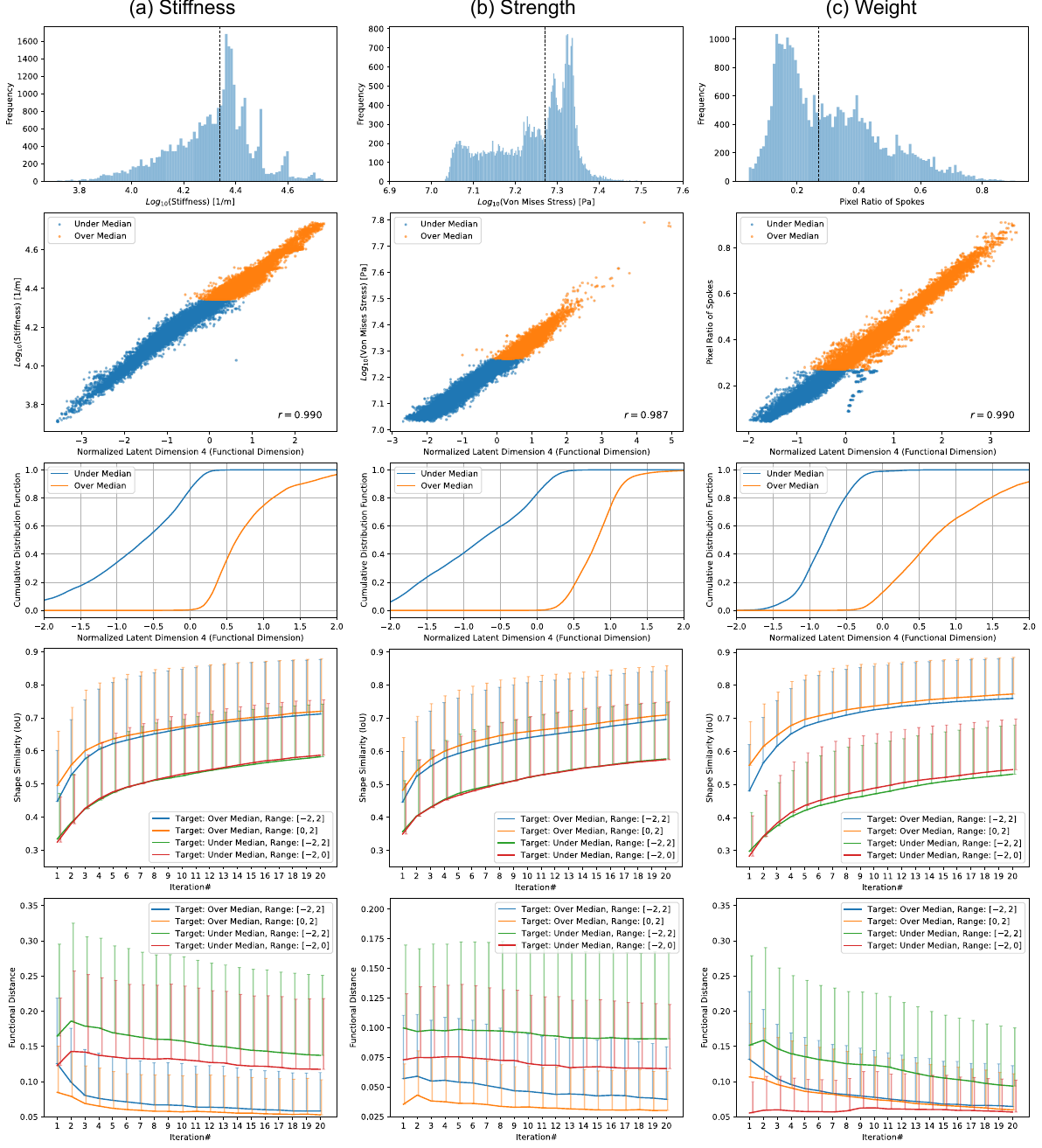}
    \caption{
        Evaluation of functional dimensions for stiffness, strength, and weight (4D).
For each functional measure, the following visualizations are shown from top to bottom: a histogram of the functional measure for the training data (the dotted line indicates the median); a scatter plot of the functional measure versus the learned functional dimension for the training data; the cumulative distribution of the training data along the learned functional dimension; the mean IoU over 20 selections (higher is better); and the mean functional distance over 20 selections (lower is better).
In the bottom two rows, error bars extend one standard deviation above
the mean; the corresponding lower error bars are omitted to reduce visual
clutter from overlapping curves.
    }
    \Description{Per-property plots showing learned functional dimensions and their effect on optimization.}
    \label{fig:effect_funcdim4}
\end{figure*}

Figures~\ref{fig:effect_funcdim3} and \ref{fig:effect_funcdim4} report the corresponding analyses for 3D and 4D latent spaces. Although the correlation coefficients in the scatter plots for (a) stiffness, (b) strength, and (c) weight (0.970, 0.977, and 0.979 for 3D; 0.990, 0.987, and 0.990 for 4D) are slightly lower than those in Figure~\ref{fig:effect_funcdim} (0.995, 0.989, and 0.996 for 5D), the same qualitative observations about design-exploration efficiency still hold.

\end{document}